\documentclass[a4paper]{jpconf}
\usepackage{graphicx}
\def\cascade{{\sc Cascade}}
\def\herwig{{\sc Herwig}}
\def\prp{\perp}
\def\kt{\ensuremath{k_\prp}}
\begin{document}
\title{Multi-jet production and 
 transverse momentum dependent parton distributions\footnote{To appear in the 
Proceedings of the EPS-HEP 2007 Conference on High Energy Physics 
(Manchester,  19-25 July 2007).}}

\author{F Hautmann$^{a}$ and H Jung$^b$}

\address{$^a$ Oxford University,  Theoretical Physics, 
Oxford OX1 3NP}

\address{$^b$ Deutsches Elektronen Synchrotron, D-22603 Hamburg}

\begin{abstract}
We study the impact  of transverse-momentum dependent 
 parton distributions on detailed features of 
  multi-jet  final states, focusing on 
 angular  jet correlations in  DIS data. 
\end{abstract}

Monte-Carlo simulations of events containing multiple  jets 
are central to a number of 
processes of interest at LHC energies, see 
e.g.~\cite{mlmhoche}, and further 
references in~\cite{heralhcproc}. These 
complex events, characterized by multiple hard scales,   
are  potentially   sensitive 
to  QCD  initial-state  radiation 
processes that depend on the large-\kt 
tail of partonic distributions and 
matrix elements~\cite{heralhcproc,jeppe06,hj_rec}. 
Such effects are  not included in  
standard shower Monte-Carlo implementing 
collinear jet evolution. 
 In this article  we report on the 
study~\cite{hj_ang} of  multi-jet  DIS final states  at HERA  
in the framework of unintegrated   parton distributions. 
Despite the much lower energy at HERA, 
we stress that,  
owing to the large phase space available  for jet production,   
the HERA data may be just as relevant as the Tevatron 
for  understanding the extrapolation to the LHC of initial-state 
radiation effects~\cite{heralhcproc}.  
We refer the reader to~\cite{hj_rec}   
for a recent discussion of  \kt-dependent  parton 
  distributions.

\begin{figure}[htb]
\vspace{50mm}
\includegraphics{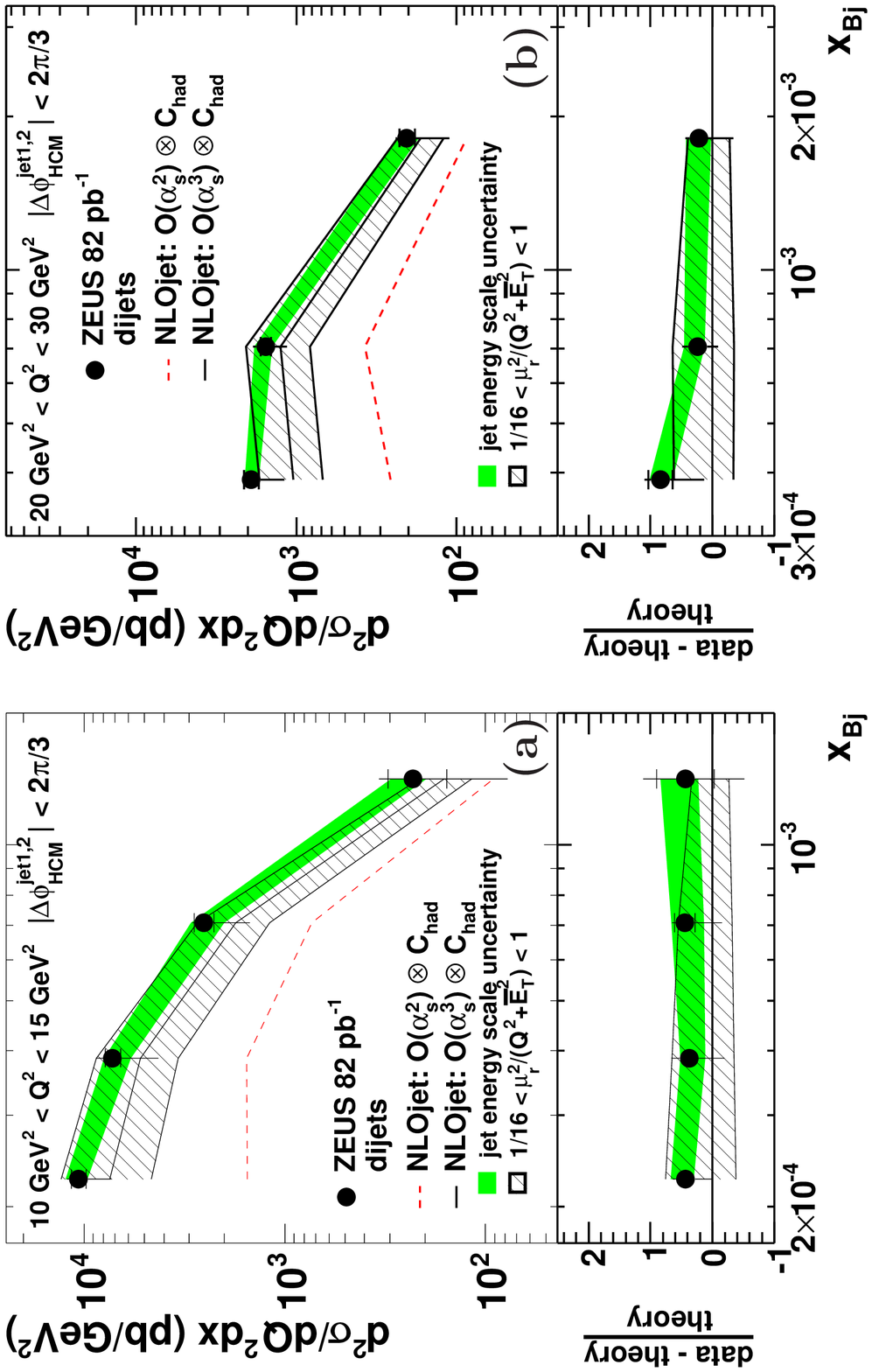}
\includegraphics{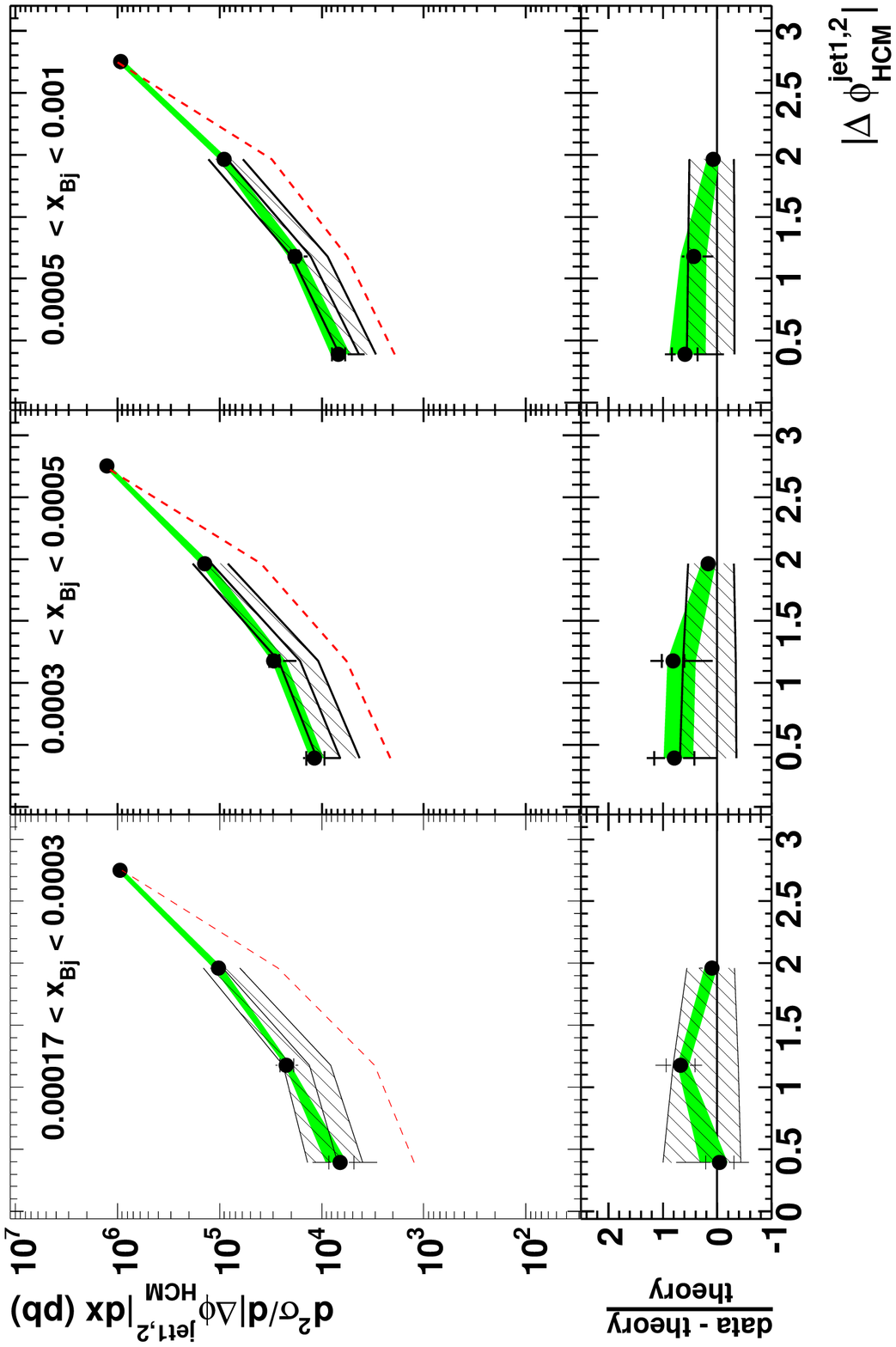}
\caption{(left) Bjorken-x  dependence and (right) azimuth dependence 
of  di-jet distributions at HERA as measured by ZEUS\protect\cite{zeus1931}.} 
\label{fig:phizeus}
\end{figure}

The ZEUS collaboration~\cite{zeus1931} 
has recently presented measurements 
of multi-jet  distributions associated  with $Q^2 > 10$ GeV$^2$ 
and Bjorken-$x$ 
as low as  $x \sim 10^{-4}$, shown in 
Fig.~\ref{fig:phizeus}~\cite{zeus1931} 
with a comparison to  
next-to-leading-order calculations~\cite{nagy}. 
The plot on the left in Fig.~\ref{fig:phizeus} shows the 
$x$-dependence of the  di-jet distribution integrated  over 
$\Delta \phi < 2 \pi / 3$, where $\Delta \phi$ is the 
azimuthal separation 
between the two high-$E_T$ 
jets. The  plot on the right 
shows the di-jet
distribution in $\Delta \phi$  for different bins of $x$.
The  data are in agreement with NLO results. 
However, the variation of the 
predictions from  order-$\alpha_s^2$ to order-$\alpha_s^3$ 
is  
very large as $x$ and $\Delta \phi$ decrease.  
The  theoretical uncertainty at 
NLO thus appears to be  sizeable, and   likely 
underestimated by the dashed error band 
 obtained from the 
conventional method of varying  the  
renormalization/factorization scale. 
The analysis~\cite{zeus1931} also indicates 
that for more 
inclusive jet cross sections 
perturbative predictions  are 
much more stable.

The jet correlations  in Fig.~\ref{fig:phizeus} of ~\cite{zeus1931}
depend on a number of   physical processes that may 
affect the stability of predictions.  
Part of these    concern the 
jet reconstruction and hadronization. In this regard 
 note that the jet algorithm  
used in~\cite{zeus1931}
is expected to have moderate 
hadronization corrections~\cite{h101} 
and to be free of nonglobal single-logarithmic  
components~\cite{nonglobjet}. 
 The kinematic cuts on the jet 
transverse momenta are set to be 
asymmetric, so as to avoid 
 double-logarithmic contributions 
in the minimum $p_T$~\cite{dasguban}. 
Nonperturbative corrections affecting  the 
jet distributions at the level of inverse powers of $Q$ 
 are expected to be 
moderate for $Q^2 > 10$ GeV$^2$.

Further, potentially important 
processes are associated with higher-order 
radiation. Higher-order contributions enhanced by 
soft and collinear gluon emission are produced from 
configurations  where  the jets are nearly 
back-to-back~\cite{delenda}. 
To the leading-log level  these contributions are 
also taken into account in    
standard shower Monte-Carlos, e.g.  \herwig~\cite{herwref}.   
However, also sizeable corrections in 
Fig.~\ref{fig:phizeus} are seen in the range of 
decreasing $\Delta \phi$, where 
the two jets are not close to back-to-back, and 
one has effectively three  
well-separated hard jets. The corrections increase as $x$ decreases. 

Motivated by this,  Ref.~\cite{hj_ang} investigates the 
possibility that 
large  higher-order terms   
come not only from the infrared emission 
but also from soft-gluon exchange 
producing relatively 
energetic multiple jets into the final state.  
In this kinematics, a sizeable \kt is built in the 
initial state, whose effects are not fully accounted for by the 
collinear approximation implemented both by 
 \herwig\ and by  
 fixed-order  calculations.  
Ref.~\cite{hj_ang} 
takes the \kt-dependence into account  both 
in the initial-state evolution and in the hard-scattering matrix 
elements, based on the 
Monte-Carlo implementation~\cite{junghgs} \cascade\  of 
high-energy factorization~\cite{hef} and  the 
fits~\cite{junghanss} to the unintegrated gluon density.

\begin{figure}[htb]
\includegraphics[width=8.2cm]{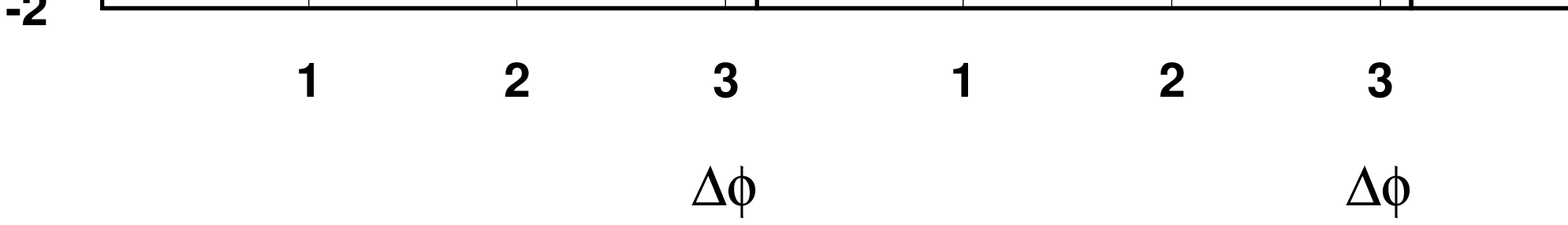}
\includegraphics[width=8.2cm]{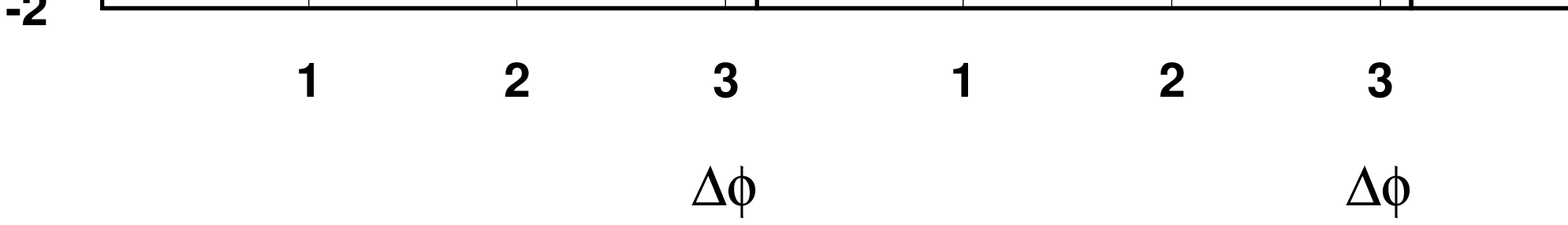}
\vspace*{-6.9cm}
\caption{Angular jet correlations~\protect\cite{hj_ang}  
obtained by \cascade\ and \herwig,  
 compared with ZEUS data~\protect\cite{zeus1931}: 
(left) di-jet cross section; (right) three-jet cross section.
} 
\label{fig:phipage}
\end{figure}

\begin{figure}[htb]
\includegraphics[width=8.2cm]{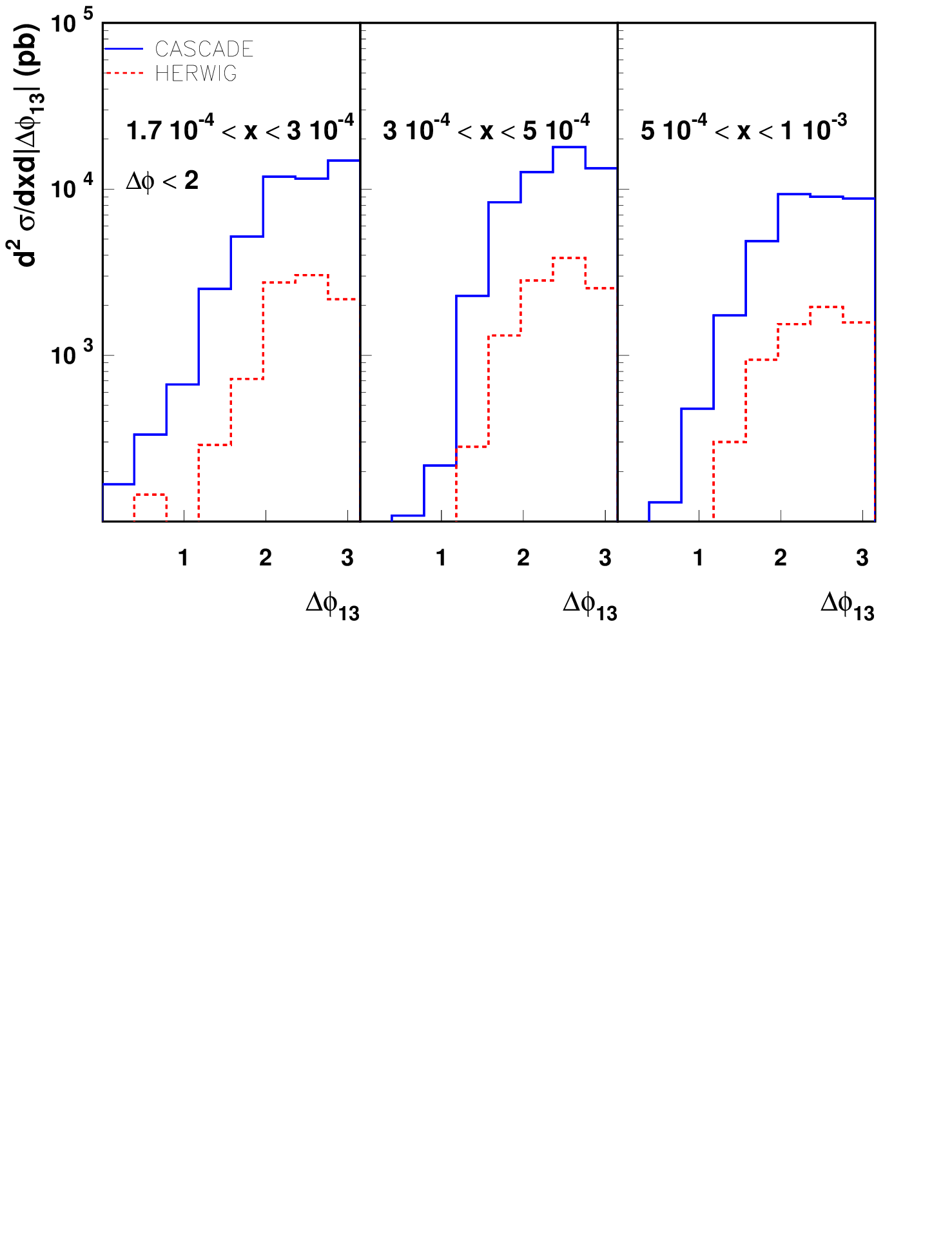}
\includegraphics[width=8.2cm]{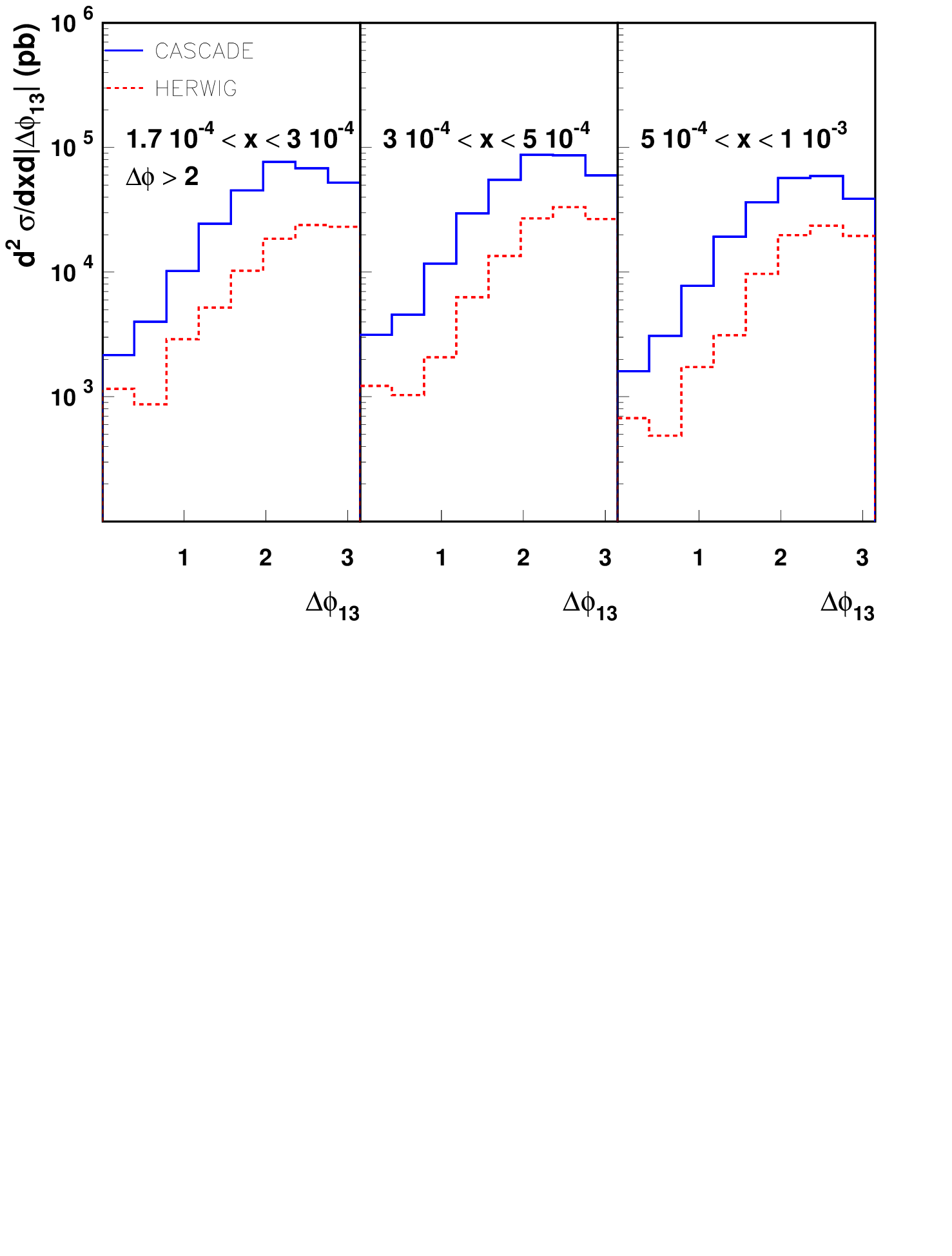}
\vspace*{-5.8cm}
\caption{Cross section in the azimuthal angle 
$\Delta \phi_{13}$ between the 
 hardest and the
3rd jet for  small
($\Delta \phi < 2$, left) and large ($\Delta \phi > 2$, right) 
 azimuthal  
separations between the leading jets~\protect\cite{hj_ang}. The \kt  
Monte-Carlo results \cascade\ are compared with \herwig .
} 
\label{fig:thirdjet}
\end{figure}

Fig.~\ref{fig:phipage} shows  results  for 
the azimuthal distribution of di-jet and three-jet 
cross sections~\cite{hj_ang}, compared with the 
measurement~\cite{zeus1931}. The description of the 
measurement by \cascade\ is
 good, whereas one can see that the collinear based parton shower
 with leading order matrix elements, as implemented by \herwig, is not
sufficient to describe the measurement in the small $\Delta \phi$ region.
\noindent 
To gain  insight into  the physical picture of the 
production process it is also worth 
looking at the angular distribution of the third jet.  
Fig.~\ref{fig:thirdjet}~\cite{hj_ang} shows 
the distribution in the azimuthal separation between  jet 1 
and  jet 3, and compares it with the result from \herwig  . 
\noindent
The \kt Monte-Carlo \cascade\  gives large differences to \herwig\
in the region where the azimuthal separations  
 $\Delta \phi$ between the 
leading jets are small  (left hand side of 
Fig.~\ref{fig:thirdjet}). It becomes closer to \herwig\ 
as  $\Delta \phi$ increases (right hand side of 
Fig.~\ref{fig:thirdjet}), consistently with the expectation 
that both Monte-Carlos give  reasonable approximations  
 in the back-to-back region.  
The behavior in Figs.~\ref{fig:phipage} and~\ref{fig:thirdjet}
is confirmed by the study of 
jet multiplicities~\cite{hj_ang}.  
The contribution of high multiplicities is found to be 
significantly increased in the \kt Monte-Carlo \cascade\ 
compared to \herwig . 
A  behavior  qualitatively similar
to that of the angular correlations is  obtained in~\cite{hj_ang} 
for the distribution in the transverse-momentum imbalance between 
the leading jets.

In conclusion, 
the analysis~\cite{zeus1931} of multi-jet production 
suggests that, 
while inclusive jet cross sections are reliably predicted 
by NLO perturbation theory, jet correlations are affected by 
sizeable higher-order corrections.  The corresponding 
theoretical uncertainties are large at NLO.  
 The results of~\cite{hj_ang} support a physical picture in which 
the large  corrections  are due not only to  contributions 
of infrared gluon 
radiation but also to the growth of the \kt in the 
initial state, enhancing effects of the $x  \ll  1 $ parton cascade. 
While the former are summed to leading-log level by the 
\herwig\ shower  (and  by the 
calculation~\cite{delenda} to the 
next-to-leading-log level), the latter require the use of 
shower Monte-Carlo implementing \kt-dependent 
parton distributions and matrix elements like \cascade  . 
The main signatures of the difference 
between the two  showerings  
are found~\cite{hj_ang} in the 
small-$\Delta \phi$ behavior  
of jet correlations and in the jet multiplicities. 
Although this analysis is for DIS, 
we note that a similar situation may be expected 
 for multi-jet final states at the LHC (unlike the Tevatron), 
owing to the large phase space becoming available  
at LHC center-of-mass energy. 

The
present \kt Monte-Carlo \cascade\ gives already 
an improved description of the
measurements compared to 
standard parton showers approaches. See e.g.~\cite{hj_rec} for a 
discussion 
of  current limitations, and  
 directions of progress, including aspects 
of the theory of unintegrated parton distributions~\cite{uop,endp}. 
Note that 
the multi-jet observables discussed in this article give 
an especially interesting probe of the approach, 
as they are much more sensitive 
than inclusive cross sections, and much less model-dependent than  
forward-region cross sections.

\end{document}